\documentclass[graybox]{svmult}

\usepackage{mathptmx}      
\usepackage{helvet}        
\usepackage{courier}     

\usepackage{makeidx}        
\usepackage{graphicx}      
                          
\usepackage{multicol}      
\usepackage[bottom]{footmisc}

\usepackage{pifont}
\newcommand{\cmark}{\ding{51}}
\newcommand{\xmark}{\ding{55}}

\usepackage{newtxtext}       
\usepackage[varvw]{newtxmath}      
\usepackage{chemformula}
\usepackage{url}

\usepackage{multirow}
\usepackage{array, booktabs}
\newcommand{\ra}[1]{\renewcommand{\arraystretch}{#1}}
\newcolumntype{P}[1]{>{\centering\arraybackslash}p{#1}}

\usepackage{siunitx}
\usepackage{titlesec}
\titleformat{\paragraph}
{\normalfont\normalsize\bfseries}{\theparagraph}{1ex}{}

\usepackage{tikz}
\usepackage{etoolbox} 
\usepackage{listofitems}

\begin{document}

\title*{Predicting Properties of Oxide Glasses Using Informed Neural Networks}

\author{Gregor Maier\inst{1,2,*} \and Jan Hamaekers\inst{1,3} \and Dominik-Sergio Martilotti\inst{1} \and Benedikt Ziebarth\inst{4}}

\institute{\inst{1}Fraunhofer Institute for Algorithms and Scientific Computing SCAI, Schloss Birlinghoven, 53757 Sankt Augustin, Germany. \\
\inst{2}Institut für Numerische Simulation, Universität Bonn, Friedrich-Hirzebruch-Allee 7,
53115 Bonn, Germany.\\
\inst{3}Fraunhofer Center for Machine Learning, Schloss Birlinghoven, 53757 Sankt Augustin, Germany.\\
\inst{4}Schott AG, Hattenbergstrasse 10, 55122 Mainz, Germany.\\
\inst{*}Corresponding author, \texttt{gregor.maier@scai.fraunhofer.de}\\}

\maketitle

\abstract{Many modern-day applications require the development of new materials with specific properties. In particular, the design of new glass compositions is of great industrial interest. Current machine learning methods for learning the composition-property relationship of glasses promise to save on expensive trial-and-error approaches. Even though quite large datasets on the composition of glasses and their properties already exist (i.e., with more than \num{350000} samples), they cover only a very small fraction of the space of all possible glass compositions. This limits the applicability of purely data-driven models for property prediction purposes and necessitates the development of models with high extrapolation power. \\
\indent In this paper, we propose a neural network model which incorporates prior scientific and expert knowledge in its learning pipeline. This informed learning approach leads to an improved extrapolation power compared to blind (uninformed) neural network models. To demonstrate this, we train our models to predict three different material properties, that is, the glass transition temperature, the Young's modulus (at room temperature), and the shear modulus of binary oxide glasses which do not contain sodium. As representatives for conventional blind neural network approaches we use five different feed-forward neural networks of varying widths and depths. \\
\indent For each property, we set up model ensembles of multiple trained models and show that, on average, our proposed informed model performs better in extrapolating the three properties of previously unseen sodium borate glass samples than all five conventional blind models.}

\section{Introduction}
\label{sec: introduction}

The development of new materials is essential for the modern-day progress in engineering applications and future-oriented technologies. Aside from ever new demands on physical and chemical materials properties, ecological issues, such as sustainability, long service life, environmental compatibility, and recyclability, are of great importance for product development in a variety of different fields. However, the common materials design process is still majorly based on the application of suitable empirical models, on past experiences and educated guesses, and on an extensive subsequent testing phase. The development of new glassy materials, in particular, would benefit to a large extent from a more resource-efficient, systematic, data-driven approach in contrast to the Edisonian trial-and-error approach which is still often used in traditional research and development~\cite{mauro2018decoding}.

The space of all possible glass compositions is very large as a glass can be made from the combination of 80 chemical elements, which leads to \num{e52} possible glass compositions~\cite{zanotto2004crystal}.
Moreover, since the influencing parameters are usually known only qualitatively or not at all, the optimization of glass material properties is inherently challenging. A trial-and-error approach to find a glass composition with specific properties for a certain application is time-consuming and often not feasible in practice. An expert-guided approach with integrating experiences from the past is usually not sufficient as well since there are interesting glass properties that are extremely difficult to predict. Especially when properties show nonlinearities, caused, for example, by the so-called borate anomaly in alkali borate glasses \cite{feller2019borate}, conventional exploration and exploitation strategies quickly reach their limits. Therefore, going beyond the area of known materials requires new approaches based on new and innovative methods. The field of machine learning (ML) provides such methods which allow to generate accurate models based on existing data in order to predict the properties of yet unseen materials. 

\subsection{Related Work}
In recent years, ML techniques have been widely used for accelerating materials design~\cite{vasudevan2019materials, wang2020machine, pilania2021machine}.
In glass science, there have been several successful attempts to use ML to predict, i.a., optical, physical, and mechanical properties of glasses~\cite{cassar2021designing, cassar2018predicting, cassar2021predicting, alcobaca2020explainable, ravinder2020deep, bishnoi2019predicting}.
Most ML models perform exceptionally well in interpolating the training data.
However, given the high-dimensional search space of all possible glass compositions and its sparse coverage by experimental data, the search for new glass materials is majorly a question of designing models which possess a high extrapolation power. We refer to~\cite{liu2021machine, ravinder2021artificial} and references therein for reviews of the current status of ML in glass science and future challenges.

To address the lack of extrapolation power, ordinary ML methods can be extended by integrating prior knowledge which exists independently of the learning task. This idea is termed \emph{informed machine learning} and we refer to the recent survey~\cite{rueden2021informed} for a taxonomy and thorough overview of its application in current ML state-of-the-art use cases. For glass design,  this idea is utilized, e.g., in~\cite{Tandia2019Machine} where the empirical MYEGA formula is integrated into a neural network architecture to predict the viscosity of a glass based on its compound fractions and temperature. In \cite{cassar2021ViscNet}, this approach is developed further by additionally integrating prior chemical and physical knowledge of the glasses' elements into the training data. Similarly, in~\cite{shih2022predicting, bishnoi2023predicting}, the authors use external chemical and physical knowledge to carefully design enriched descriptors of glass compositions which are used as inputs for ML models to predict properties of oxide glasses. In~\cite{krishnan2018predicting} and~\cite{liu2019predicting}, the authors predict the dissolution kinetics of silicate glasses in an informed manner by suitably splitting the training data and using a descriptor which encodes the glasses' network structure, respectively. They demonstrate the superior performance of the informed approach compared to the uninformed approach. This superiority is also shown in~\cite{bodker2022predicting}, where the authors design a neural network model which is informed by statistical mechanics in order to predict structural properties of oxide glasses.

\subsection{Contributions}
In this paper, we propose a new ML model based on neural networks for the property prediction of oxide glasses which integrates prior knowledge in order to achieve a high degree of extrapolation of the training data. We modify the ideas from~\cite{cassar2021ViscNet} in order to predict three material properties, that is, the \emph{glass transition temperature} \(T_g\), the \emph{Young's modulus \(E\) (at room temperature)}, and the \emph{shear modulus} \(G\). We focus our analysis on binary oxide glasses, that is, oxide glasses which consist of exactly two compounds. Our model is informed in the sense that we explicitly integrate prior knowledge into the design of our training data, the hypothesis set, and the final hypothesis at four major points in our learning pipeline. We place emphasis on explaining how this is done in detail in terms of the taxonomy in~\cite{rueden2021informed}. Especially the design of the network architecture to realize permutation invariance with respect to the input features seems, to the best of our knowledge, to be new in the field of glass materials modeling. 

To examine the extrapolation power of our models, we train and validate them on glass samples which do not contain sodium in their compositions. The trained models are then used to predict the properties of sodium borate glass compositions with varying compound fractions. For each property, we train multiple models and study the average performance of the model ensemble.
To demonstrate the superiority of the informed model ensemble compared to blind (uninformed) approaches, we perform the same experiments with five standard fully connected feed-forward neural networks of varying widths and depths without integration of any prior knowledge. We compare the results quantitatively in terms of error metrics and qualitatively in terms of a meaningful approximation of the respective composition-property curves. \\

\noindent\textbf{Outline.} \hspace{1ex}
The remainder of this paper is organized as follows: In Sect.~\ref{sec: methodology}, we explain our methodology. That is, in Sect.~\ref{subsec: data collection and preparation}, we present our automated pipeline for collecting and preparing data for model training, validation, and testing. In Sect.~\ref{subsec: model setups}, we describe the different model setups in the blind and the informed setting. In Sect.~\ref{subsec: model training and evaluation}, we explain how we train and evaluate our models. We discuss the results of our experiments in Sect.~\ref{sec: results and discussion} and conclude our findings in Sect.~\ref{sec: conclusion and outlook}. \\

\noindent\textbf{Notation.} \hspace{1ex}
For notational convenience, we use the letter \Vec{P} whenever we refer to one of the three properties \(T_g\), \(E\), or \(G\). Moreover, for all entities which exist for every property \Vec{P}, we use the prefix ``\Vec{P}-'' to specify the respective entity. For example, given \Vec{P}, we refer to the dataset that is used to train a model for predicting \Vec{P} by ``\Vec{P}-training set''.

Moreover, we use the symbols \(\mathbb{N}\) and \(\mathbb{R}\) to denote the set of positive integers and the set of real numbers, respectively.

\section{Methodology}
\label{sec: methodology}

The prediction quality of any data-driven machine learning algorithm in the context of supervised learning is strongly dependent on the quantity and quality of the training data. Before presenting our neural network approach to the problem of glass property prediction in detail in Sect.~\ref{subsec: model setups} and Sect.~\ref{subsec: model training and evaluation}, we therefore describe in the following Sect.~\ref{subsec: data collection and preparation} how we collect and prepare our data.

\subsection{Data Collection and Preparation}
\label{subsec: data collection and preparation}

We use data from the INTERGLAD Ver. 8 database~\cite{interglad} and the SciGlass database \cite{sciglass} and merge them together into a common \texttt{glassmodel} database. For the identification of oxide glasses we follow the same definition as in~\cite{alcobaca2020explainable} and only consider glasses whose mole atomic fraction of oxygen is at least \num{0.3} and whose compounds do not contain the chemical elements S, H, C, Pt, Au, F, Cl, N, Br, and~I, which could affect the balance of oxygen. The resulting \texttt{glassmodel} database of oxide glasses consists of \num{420973} glass  samples in total. It lists the mole atomic fractions of 118 chemical elements and the mole fractions of 439 compounds, i.e., the oxides that a glass composition consists of, together with the values of \num{87} material properties. However, among the 118 elements, only 66 elements appear with non-vanishing fraction in at least one glass sample. Among the 439 compounds, only 183 compounds appear with non-vanishing fraction in at least one glass sample.

\begin{figure}[h]
\centering
\sidecaption
\includegraphics[scale=0.5,trim={4cm 2.2cm 3cm 2cm},clip]{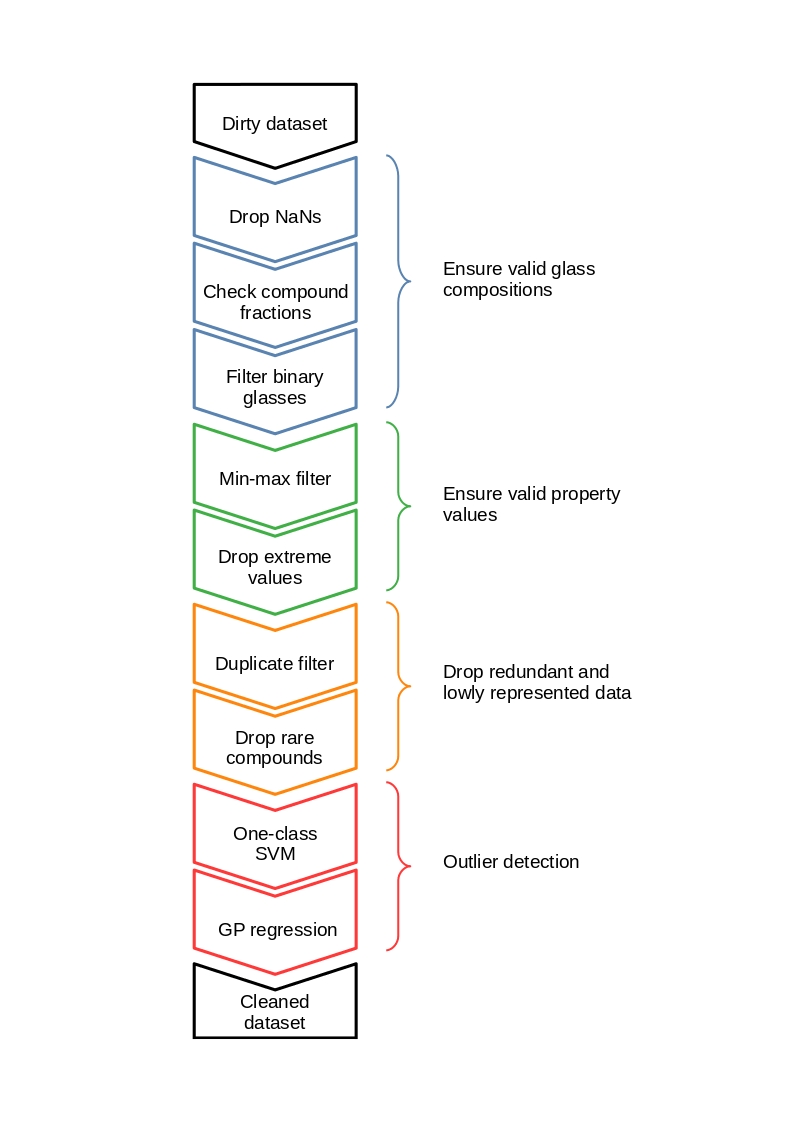}
\caption{Steps in the preprocessing pipeline as described in Sect.~\ref{subsec: data collection and preparation}.}
\label{fig: preprocessing pipeline}
\end{figure}

To obtain clean data for training, validating, and testing our models, we apply a sequence of preprocessing steps which follows in parts the procedure described in~\cite{alcobaca2020explainable, cassar2021designing}. For each glass property \Vec{P}, we extract clean data from the ``dirty'' \texttt{glassmodel} database in an automated fashion in form of a preprocessing pipeline whose steps are schematically depicted in Figure~\ref{fig: preprocessing pipeline}. The number of samples which are dropped in each step is shown in Table~\ref{table: preprocessing info}.

We begin with all samples from the entire \texttt{glassmodel} database. As a first step, we make sure that all glass samples have numerically valid entries. That is, we first remove glass samples which have a Not-a-Number (NaN) entry for at least one compound fraction. Moreover, we drop all glasses which have NaN entries for \Vec{P}.\footnote{At the end of this and all the following preprocessing steps, we always drop all compounds which do not appear in any of the glass samples that are present in the dataset at the respective preprocessing stage.}

Next, we make sure that all glass samples are physically valid binary glass compositions.
For this, we first discard glasses whose compound fractions do not add up to a value in the closed range between \(0.9999\) and \(1.0001\). Then, we exclude all samples which do not consist of exactly two compounds.

To ensure physically valid property values, we fix a closed range of values between a minimum and maximum cut-off value for each property \Vec{P} (see Table \ref{table: min_max, duplicate_thresholds}). We determine these values by investigating the distribution of the glass samples with respect to their \Vec{P}-values in the datasets that result from the preprocessing pipeline up to this point. Property values outside of this range are considered non-physical but may be present in the database due to typos or other mistakes. Hence, we drop each glass sample with a \Vec{P}-value outside of the respective range. 

As the minimum and maximum values are rather crude bounds, in a further step, we remove glasses with extreme \Vec{P}-values, which have a high chance of still appearing in the datasets again because of typos or other mistakes. To do so, we compute the \(0.05\)th percentile and the \(99.95\)th percentile of \Vec{P}-values among all remaining glass samples and subsequently discard all glasses with \Vec{P}-values below the lower percentile or above the upper percentile.

A lot of glass samples appear in both the INTERGLAD and the SciGlass database. Consequently, there may be many duplicates among the remaining data points at this stage of the preprocessing pipeline. We therefore apply a duplicate filter which consists of the following steps:
\begin{enumerate}
    \item We group all glasses with the same (up to the fifth decimal place) compound fractions.
    \item For each such group we do the following:
    \begin{enumerate}
        \item [2.1] We drop all but the first sample which agree \emph{exactly} in their values of \Vec{P}.
        \item [2.2] We compute the midpoint of the range of values of \Vec{P} among all remaining samples.
        \item [2.3] If the \Vec{P}-value of every sample has a distance to the midpoint smaller than a certain \Vec{P}-dependent threshold (see Table \ref{table: min_max, duplicate_thresholds}), then, as a representative of the group of duplicates, we select the first sample in the group, assign to it the median of the \Vec{P}-values of the samples in the group, and drop all other samples.
        Otherwise, we discard the whole group of glass samples.
        
        The values for the duplicate thresholds are determined by using domain know-ledge and investigating the average spread of \Vec{P}-values in a group of duplicates.
    \end{enumerate}
\end{enumerate}

In the next step, we deal with compounds of low representability and iteratively drop compounds which appear in less than one percent of all remaining glass samples. This allows us to reduce the dimension of the compound space and leaves us only with glasses whose compounds are present in sufficiently many samples in order to use them for robust model training.\footnote{At the end of each iteration, we again drop those samples whose compound fractions do not add up to a value in the closed range between \(0.9999\) and \(1.0001\).}

As a final step, we apply an outlier detection based on a one-class support vector machine (SVM) followed by an outlier detection based on Gaussian process (GP) regression.\footnote{We fit a Gaussian process to the data and drop samples with a too large deviation in their \Vec{P}-value from the respective mean curve.}

The resulting cleaned datasets encompass all problem-specific information which is available for each glass property \Vec{P}. Given \Vec{P}, we denote the elements and compounds which are present (with non-vanishing value in at least one glass sample) in the corresponding cleaned dataset as \Vec{P}-elements and \Vec{P}-compounds, respectively. The cleaned datasets are subsequently split into training, validation, and test sets as described in Sect.~\ref{subsec: model training and evaluation}.

\begin{table*}[h!]\centering
\caption{Data reduction in each step of the preprocessing pipeline. The first and last row show the number of samples (\#S) and number of compounds (\#C) which are present in the dirty and cleaned dataset, respectively. The rows in between show the number of samples and compounds which are dropped in the respective preprocessing steps. We remark that in the cleaned datasets, the number of \emph{elements} appearing with non-vanishing fraction in at least one glass sample is 32 for \(T_g\), 23 for \(E\), and \(24\) for \(G\) (and therefore coincides with \#C).}
\label{table: preprocessing info}
\ra{1.3}
    \begin{tabular}{@{} p{3.5cm} S[table-format=7.0] S[table-format=4.0] S[table-format=7.0] S[table-format=4.0] S[table-format=7.0] S[table-format=4.0] @{}}
        \hline\noalign{\smallskip}
        \Vec{P} & \multicolumn{2}{r}{\(T_g\)} & \multicolumn{2}{r}{\(E\)} & \multicolumn{2}{r}{\(G\)} \\
         & \multicolumn{1}{r}{\#S} & \multicolumn{1}{r}{\#C} & \multicolumn{1}{r}{\#S} & \multicolumn{1}{r}{\#C} & \multicolumn{1}{r}{\#S} & \multicolumn{1}{r}{\#C} \\
        \noalign{\smallskip}\svhline\noalign{\smallskip}
        Dirty dataset & 420973 & 439 & 420973 & 439 & 420973 & 439 \\
        \noalign{\smallskip}\hline\noalign{\smallskip}
        Drop NaNs & 344247 & 283 & 396460 & 329 & 410589 & 356 \\
        Check compound fractions & 17483 & 1 & 5482 & 0 & 1867 & 0 \\
        Filter binary glasses & 50681 & 83 & 16651 & 57 & 6386 & 38 \\
        Min-max filter & 16 & 0 & 21 & 0 & 15 & 0 \\
        Drop extreme values & 10 & 0 & 4 & 0 & 4 & 0 \\
        Duplicate filter & 5902 & 0 & 1577 & 0 & 1253 & 2 \\
        Drop rare compounds & 229 & 40 & 79 & 30 & 67 & 19 \\
        One-class SVM & 22 & 0 & 5 & 0 & 10 & 0 \\
        GP regression & 205 & 0 & 62 & 0 & 63 & 0 \\
        \noalign{\smallskip}\hline\noalign{\smallskip}
        Cleaned dataset & 2178 & 32 & 632 & 23 & 719 & 24\\
        \noalign{\smallskip}\hline\noalign{\smallskip}
    \end{tabular}
\end{table*}

\begin{table*}[h!]\centering
\caption{Minimum and maximum cut-off values and duplicate thresholds used in the preprocessing pipeline.}
\label{table: min_max, duplicate_thresholds}
\ra{1.3}
    \begin{tabular}{@{} p{0.3cm} p{1cm} S[table-format=2.2] S[table-format=1.1] S[table-format=1.2] @{}}
        \hline\noalign{\smallskip}
        \Vec{P} & & {\ Min. cut-off \ } & {\ Max. cut-off \ } & {\ Duplicate threshold \ } \\
        \noalign{\smallskip}\svhline\noalign{\smallskip}
        \(T_g\) & (\textdegree{}C) & 50 & \num{1.8e3} & 5.0 \\
        \(E\) & (GPa) & 5.0 & \num{2.0e2} & 1.5\\
        \(G\) & (GPa) & 0.10 & \num{2.0e2} & 0.75\\
        \noalign{\smallskip}\hline\noalign{\smallskip}
    \end{tabular}
\end{table*}

\subsection{Model Setups}
\label{subsec: model setups}

We use neural networks for the approximation of the composition-property relationship of binary oxide glasses. The target quantity is given by one of the respective properties \Vec{P}. The composition of a glass can be represented in various ways. Designing a representation in form of a feature vector that encodes a given glass composition in a way that is suitable as input for a neural network is an essential part of the modeling process and is one of the key differences between the blind (uninformed) and our informed learning approach. A second difference lies in the design of the model architecture where the black-box modeling approach of standard blind feed-forward neural networks can be leveraged in the informed setting by integrating prior scientific knowledge.

In the following Sect.~\ref{subsubsec: blind models} and Sect.~\ref{subsubsec: informed model}, we describe in detail the choice of the feature vectors and the network architectures for the blind and informed models, respectively, and highlight their differences.

\subsubsection{Blind Models}
\label{subsubsec: blind models}

In the \emph{blind} approach, only the available problem-specific data is used to design a suitable ML model for the composition-property relationship of oxide glasses. This approach is blind or uninformed in the sense that no prior knowledge that exists independently of the learning task is integrated into the model setup.

\paragraph{Feature Vectors}
\label{subsubsubsec: blind: feature vectors}

Each glass composition is, by definition, uniquely determined by its compound fractions. It is therefore natural to use the compound fractions, grouped together in a feature vector for a given glass composition, as input for a neural network model to predict one of the glass's properties.

\paragraph{Network Architectures}
\label{subsubsubsec: blind: network architectures}

If no further information is available, the standard architectural design of a neural network is given by a (fully connected) feed-forward neural network (FFNN)~\cite{goodfellow2016ml}. This class of models satisfies the universal approximation theorem, that is, for any continuous function on a compact domain there exists a FFNN which approximates the function within a given arbitrary tolerance \cite{cybenko1989approximation, hornik1991approximation, pinkus1999approximation}.
This result justifies the usage of the set of FFNNs as hypothesis space. In the context of glass materials research, this approach is followed for example in~\cite{cassar2018predicting, ravinder2020deep} to model several different properties of oxide glasses.

A fully connected FFNN is characterized by (i) its input and output dimensions, i.e., the number of units in its input and output layer, respectively, (ii) its depth, i.e., the number of layers (without counting the input layer), and (iii) the width, i.e., the number of units, of each hidden layer. An example architecture of a fully connected FFNN is shown in Fig.~\ref{fig: FFNN}.

\tikzset{>=latex} 
\colorlet{myred}{red!80!black}
\colorlet{myblue}{blue!80!black}
\colorlet{mygreen}{green!60!black}
\colorlet{mydarkred}{myred!40!black}
\colorlet{mydarkblue}{myblue!40!black}
\colorlet{mydarkgreen}{mygreen!40!black}
\tikzstyle{node}=[very thick,circle,draw=myblue,minimum size=22,inner sep=0.5,outer sep=0.6]
\tikzstyle{connect}=[->,thick,mydarkblue,shorten >=1]
\tikzset{
  node 1/.style={node,mydarkgreen,draw=mygreen,fill=mygreen!25},
  node 2/.style={node,mydarkblue,draw=myblue,fill=myblue!20},
  node 3/.style={node,mydarkred,draw=myred,fill=myred!20},
}
\def\nstyle{int(\lay<\Nnodlen?min(2,\lay):3)}

\begin{figure}[h!]
\centering
    \begin{tikzpicture}[x=2cm,y=1cm]
      \readlist\Nnod{3,4,4,4,2} 
      \readlist\Cstr{x,h^{(\prev)},y} 
      
      \foreachitem \N \in \Nnod{
        \def\lay{\Ncnt}
        \pgfmathsetmacro\prev{int(\Ncnt-1)} 
        \foreach \i [evaluate={\c=int(\i==\N); \y=\N/2-\i;
                     \x=\lay; \n=\nstyle;
                    }] in {1,...,\N}{
          \node[node \n] (N\lay-\i) at (\x,\y) {};
          
          \ifnumcomp{\lay}{>}{1}{ 
            \foreach \j in {1,...,\Nnod[\prev]}{
              \draw[white,line width=1.2,shorten >=1] (N\prev-\j) -- (N\lay-\i);
              \draw[connect] (N\prev-\j) -- (N\lay-\i);
            }
            \ifnum \lay=\Nnodlen
              \draw[connect] (N\lay-\i) --++ (0.5,0); 
            \fi
          }{
            \draw[connect] (0.5,\y) -- (N\lay-\i);
          }
        }
      }
      
      \node[above=3,align=center,mydarkgreen] at (N1-1.90) {Input\\[-0.2em]layer};
      \node[above=2,align=center,mydarkblue] at (N2-1.90) {Hidden\\[-0.2em]layer 1};
      \node[above=2,align=center,mydarkblue] at (N3-1.90) {Hidden\\[-0.2em]layer 2};
      \node[above=2,align=center,mydarkblue] at (N4-1.90) {Hidden\\[-0.2em]layer 3};
      \node[above=3,align=center,mydarkred] at (N\Nnodlen-1.90) {Output\\[-0.2em]layer};
    \end{tikzpicture}
    
\caption{Schematic architecture of a fully connected FFNN with input dimension \(3\), output dimension \(2\), depth \(4\), and constant width \(4\). Each layer represents an affine linear function. The additive bias nodes are not shown. In case of the blind models, the input nodes store the compound fractions of a given glass sample and the output node provides the predicted value for \Vec{P}. Image adapted from~\cite{NNtikz}.}
\label{fig: FFNN}
\end{figure}
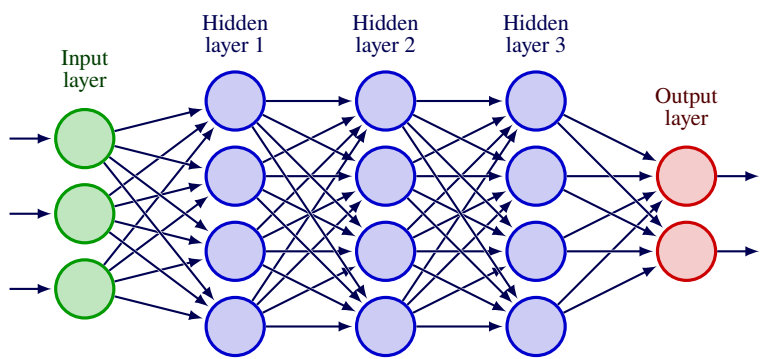

We use a variety of different FFNNs as benchmark models which we compare our informed model to. To capture the main architectural trends of designing a FFNN and their effects on the prediction quality, we consider five different FFNNs with depths \(L = 2, 4, 8, 16, 32\) for each property \Vec{P}. The input dimensions are determined by the number of respective \Vec{P}-compounds and the output dimension is always one as we predict scalar-valued properties.
For each \Vec{P}-model, we choose the width to be constant for all hidden layers such that the total number of trainable parameters is roughly the same among all \Vec{P}-models including the informed model which we describe in Sect.~\ref{subsubsec: informed model}. Each hidden layer is a linear layer with an additive bias term. In accordance to the universal approximation theorem, the output layer is a linear layer with no additive bias term. The exact dimensions of all models are summarized in Table \ref{table: model hyperparameters}. In all models, we use the rectified linear unit (ReLU) as activation function.

\begin{table}[h]\centering
\ra{1.3}
\caption{Model hyperparameters}
\label{table: model hyperparameters}
    \begin{tabular}{@{} p{1.8cm} S[table-format = 5.0] S[table-format = 5.0] S[table-format = 5.0] S[table-format = 5.0] S[table-format = 5.0] S[table-format = 5.0] S[table-format = 5.0] S[table-format = 5.0] @{}}
    \hline\noalign{\smallskip}
    \multirow{2}{*}{\(T_g\)} & \multicolumn{5}{c}{\multirow{2}{*}{Blind}} & \multicolumn{3}{c}{Informed} \\ 
    \cline{7-9} 
     & \multicolumn{5}{c}{} & {Former} & {Non-former} & {Down} \\ 
    \noalign{\smallskip}\svhline\noalign{\smallskip}
    Input dim. & 32 & 32 & 32 & 32 & 32 & 28 & 28 & 32 \\ 
    Length \(L\) & 2 & 4 & 8 & 16 & 32 & 4 & 4 & 4 \\
    Width \(W\) & 304 & 63 & 38 & 26 & 18 & 32 & 32 & 32 \\
    Output dim. & 1 & 1 & 1 & 1 & 1 & 14 & 18 & 1 \\ 
    \cline{7-9}
    \#Parameters & 10336 & 10206 & 10184 & 10712 & 10872 & \multicolumn{3}{c}{\num{10336}} \\ \noalign{\smallskip}\hline\noalign{\smallskip}
    \\ 
    \hline\noalign{\smallskip}
    \multirow{2}{*}{\(E\)} & \multicolumn{5}{c}{\multirow{2}{*}{Blind}} & \multicolumn{3}{c}{Informed} \\ 
    \cline{7-9} 
     & \multicolumn{5}{c}{} & {Former} & {Non-former} & {Down} \\ 
    \noalign{\smallskip}\svhline\noalign{\smallskip}
    Input dim. & 23 & 23 & 23 & 23 & 23 & 26 & 26 & 23 \\ 
    Length \(L\) & 2 & 4 & 8 & 16 & 32 & 4 & 4 & 4 \\ 
    Width \(W\) & 385 & 63 & 38 & 25 & 17 & 32 & 32 & 32 \\ 
    Output dim. & 1 & 1 & 1 & 1 & 1 & 10 & 13 & 1 \\ 
    \cline{7-9}
    \#Parameters & 9625 & 9639 & 9842 & 9725 & 9605 & \multicolumn{3}{c}{\num{9623}} \\ 
    \noalign{\smallskip}\hline\noalign{\smallskip}
    \\ 
    \hline\noalign{\smallskip}
    \multirow{2}{*}{\(G\)} & \multicolumn{5}{c}{\multirow{2}{*}{Blind}} & \multicolumn{3}{c}{Informed} \\ 
    \cline{7-9} 
     & \multicolumn{5}{c}{} & {Former} & {Non-former} & {Down} \\ 
    \noalign{\smallskip}\svhline\noalign{\smallskip}
    Input dim. & 24 & 24 & 24 & 24 & 24 & 26 & 26 & 24 \\ 
    Length \(L\) & 2 & 4 & 8 & 16 & 32 & 4 & 4 & 4 \\ 
    Width \(W\) & 372 & 63 & 38 & 25 & 17 & 32 & 32 & 32 \\ 
    Output dim. & 1 & 1 & 1 & 1 & 1 & 10 & 14 & 1 \\ 
    \cline{7-9}
    \#Parameters & 9672 & 9702 & 9880 & 9750 & 9622 & \multicolumn{3}{c}{\num{9688}} \\ 
    \noalign{\smallskip}\hline\noalign{\smallskip}
    \end{tabular}
\end{table}

\subsubsection{Informed Model}
\label{subsubsec: informed model}

Rather than just using the compound fractions as input features for a neural network, we can increase the informational capacity of a glass sample's representation by utilizing characteristic chemical and physical quantities of each element which is present in the given glass sample and provide them as additional inputs to a neural network. Features which are carefully engineered in such an informed manner can lead to an improved prediction quality of the model, given that the model's expressive power is large enough. The latter issue is a question of the model's architecture. If there are too few parameters, the model will underfit the training data no matter how carefully we designed the input features. If there are too many parameters, however, the model might overfit the training data and pick up on spurious patterns and noise in the input features. In general, by building as much prior information as possible into the model's architecture we expect to obtain a more robust inference behavior, especially in the extrapolation regime.

\paragraph{Feature Vectors}
\label{subsubsubsec: informed: feature vectors}

Compared to the uninformed approach, we change our viewpoint and identify a given glass composition not by the fractions of its compounds but by the mole atomic fractions of its elements. For each element, there is additional extensive scientific knowledge about its chemical and physical properties, which exists independently of our learning problem. According to the taxonomy in~\cite{rueden2021informed} this knowledge is represented as a weighted graph. Its nodes are given by elements and properties and each element node is connected with a property node via an edge which is weighted by the element's respective property value. We integrate this knowledge into our training data by designing feature vectors in a hybrid fashion. We partly follow the approach used in~\cite{cassar2021ViscNet}, where, for each element, the authors extract physical and chemical properties, i.a., from the Python library \texttt{mendeleev}~\cite{mendeleev2014}. They design and select feature vectors for a neural network model in a way that allows them to complement information from the space of chemical compositions with information from the space of chemical and physical properties.

In our case, we first extract for each glass property \Vec{P} and each \Vec{P}-element a list of characteristic chemical and physical properties from \texttt{mendeleev}. We drop properties with non-numeric values and only keep those which are available for all \Vec{P}-elements. We also drop properties which we consider to be unrelated to the elements' influence on the glass material properties, namely, the elements' \emph{abundances in the earth crust} and the elements' \emph{dipole polarizability uncertainties}.
We refer to~\cite{mendeleev2014} and references therein for a detailed explanation of all the available properties in the \texttt{mendeleev} library.

Among the remaining properties, we drop those which are highly correlated. More specifically, we compute the standard pairwise Pearson correlation coefficients and iteratively, for each pair of properties whose coefficient is larger than \(0.95\), we only keep one of the two properties. The resulting list of properties is given in Table \ref{table: mendeleev properties}.

For each glass property \Vec{P} and each \Vec{P}-element, we group the resulting element properties into vectors, which we call \Vec{P}-element-vectors. These element-vectors represent all prior knowledge about the elements which exists independently of the learning problem. 

In the next step, we complement this information with our given problem-specific data. For each glass property \Vec{P} and a given glass sample, we consider the collection of \Vec{P}-element-vectors corresponding to the sample's elements and extend each of them by one more component which lists the mole atomic fraction of the respective element in the glass sample. Eventually, we obtain for each glass sample a collection of feature vectors \((v_1,\dots,v_M)\in\mathbb{R}^{d\times M}\), where we arrange the tuple in lexicographical order of the elements' symbols. Here, \(M = M(\Vec{P})\) is the number of \Vec{P}-elements and \(d = d(\Vec{P})\) is the number of properties resulting from the property extraction process described above (including the entry with the mole atomic fraction). Each glass sample can thus be represented as a point in a subspace \(\Omega = \Omega(\Vec{P})\subset \mathbb{R}^{d\times M}\).

\paragraph{Network Architecture}
\label{subsubsubsec: informed: network architecture}

For each glass property \Vec{P}, we want to design a neural network which approximates the functional relationship \(f: \Omega \to \mathbb{R}, \Vec{V}\mapsto \Vec{P}(\Vec{V})\), where \(\Vec{P}(\Vec{V})\) is the value of property \Vec{P} for the glass sample with representation \(\Vec{V} = (v_1,\dots,v_M)\).
We design the architecture of the network by two leading principles in the spirit of informed learning.

First, we observe that the order in which the feature vectors are passed to the function \(f\) actually does not matter. That is, the function \(f\) is \emph{permutation invariant} with respect to the order of the input vectors. More specifically, \(f\) is a function on sets of the form \(\{v_1,\dots, v_M\}\). In terms of the taxonomy in~\cite{rueden2021informed} we use this scientific knowledge, which is represented as a spatial invariance, and directly integrate it into the architecture of the network which we use to approximate \(f\). It is shown in \cite{zaheer2017deep} that such a function \(f\) on sets can be written as
\begin{equation}\label{eq: func on sets}
    f(\{v_1, \ldots, v_M\}) =  \psi\left(\sum_{i=1}^M \phi(v_i)\right) \, ,
\end{equation}
where \(\phi : \mathbb{R}^d \to \mathbb{R}^N\) denotes an inner embedding function with \(N\in\mathbb{N}\) being an appropriately chosen embedding dimension, and \(\psi : \mathbb{R}^N \to \mathbb{R}\) denotes an outer (downstream) function. Here, \(\phi\) and \(\psi\) can be approximated by neural networks. Using the universal approximation theorem of neural networks, the right-hand-side of \eqref{eq: func on sets} yields architectures of neural networks which, in principal, can approximate \(f\) arbitrarily well.

In our specific use case, we can refine the network's architecture even further by integrating prior chemical knowledge. Glass oxides can be categorized in three groups~\cite{boubata2013thermodynamic}. \emph{Glass formers} are oxides that can readily form a glassy material and build the backbone of a glass's network structure. \emph{Glass modifiers} are oxides that cannot form a glassy material by themselves but influence its material properties when mixed with a glass former. \emph{Glass intermediates} are oxides which can act both as a glass former as well as a glass modifier depending on the respective cation's oxidation number. For our purposes, we only differentiate between oxides which are glass formers and oxides which are not glass formers. We refer to the latter group as \emph{glass non-formers}.
We use the classification proposed in~\cite{boubata2013thermodynamic} to determine for every element whether its oxide is a glass former or a glass non-former. The classification is shown in Table~\ref{table: formers and non-formers}. 

\begin{table}[h]\centering
\caption{Classification of elements based on the glass-forming and glass-non-forming properties of their oxides. The last row shows all elements whose oxides are glass formers according to the classification in~\cite{boubata2013thermodynamic}. The second column lists, for given \Vec{P}, the respective \Vec{P}-elements whose oxides are glass formers and which are a subset of the elements in the last row. The fourth column lists all respective \Vec{P}-elements whose oxides are glass non-formers. We classify oxygen as a glass non-former.}
\label{table: formers and non-formers}
\ra{1.3}
    \begin{tabular}{@{} p{1cm}p{3cm}P{2cm}p{3cm}P{1.8cm} @{}} 
    \hline\noalign{\smallskip}
    \Vec{P} & Formers & \#Formers & Non-formers & \#Non-formers \\ 
    \noalign{\smallskip}\svhline\noalign{\smallskip}
    \(T_g\) & As, B, Bi, Ge, Mo, P, Pb, Sb, Si, Sn, Te, Tl, V, W & 14 & Ag, Al, Ba, Ca, Cs, Cu, Fe, Ga, K, La, Li, Mg, Na, O, Rb, Sr, Ti, Zn & 18 \\
    \(E\) & B, Bi, Ge, Mo, Nb, P, Pb, Si, Te, V & 10 & Al, Ba, Ca, Co, Cs, K, Li, Mg, Na, O, Sr, Ti, Zn & 13 \\
    \(G\) & B, Bi, Ge, Mo, Nb, P, Pb, Si, Te, V & 10 & Al, Ba, Ca, Co, Cs, K, Li, Mg, Na, O, Rb, Sr, Ti, Zn & 14 \\ 
    \noalign{\smallskip}\hline\noalign{\smallskip}
    All & As, B, Bi, Ge, Mo, Nb, P, Pb, Sb, Se, Si, Sn, Ta, Te, Tl, V, W &  17 &  &  \\ 
    \noalign{\smallskip}\hline\noalign{\smallskip}
    \end{tabular}
\end{table}

The scientific knowledge whether an element's oxide has glass-forming or glass-non-forming ability is naturally represented as a simple knowledge graph, where each element is represented by a node. There is also a glass former node and a glass non-former node. Each element node is connected via an edge with the glass former node or the glass non-former node depending on whether the element's oxide is a glass former or a glass non-former. Due to the largely different influence on a glass's properties, we integrate this prior knowledge additionally into our hypothesis set by using two functions to treat glass formers and non-formers separately. The \emph{glass former network} receives as input only feature vectors of elements whose oxides are glass formers. The \emph{glass non-former network} receives as input all other elements whose oxides, by definition, are glass non-formers. The outputs of the glass former network are added together, as are the outputs of the glass non-former network. The results are concatenated and then used as input for the \emph{downstream network} which yields the final prediction for the respective property \Vec{P}.

\begin{figure}[h]
\centering
\includegraphics[scale=.23]{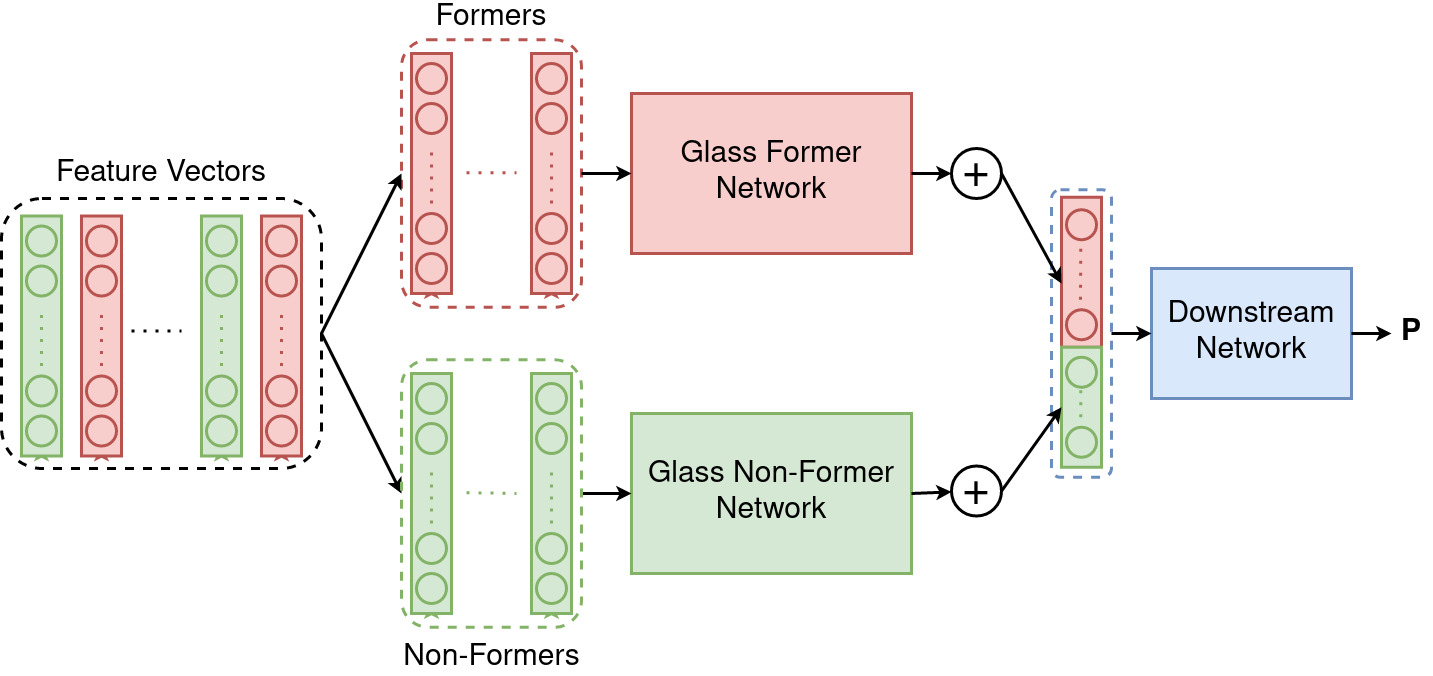}
\caption{Architecture of the informed model. The feature vectors of the elements are split according to whether the elements' oxides are glass formers or glass non-formers and input to separate neural networks. The results of the latter are first summed individually and then concatenated to a vector which is used as input for the final downstream neural network that predicts a value for property~\Vec{P}.}
\label{fig: InformedNet}
\end{figure}

More specifically, let \(\Omega = \Omega_f \cup \Omega_{nf}\) be the decomposition of \(\Omega\) into the space \(\Omega_f\) of representations of glass formers and the space \(\Omega_{nf}\) of representations of glass non-formers. We then replace the inner function \(\phi\) in~\eqref{eq: func on sets} by two separate functions, \(\phi_{f}: \Omega_f \to \mathbb{R}^{N_f}\), \(N_f\in\mathbb{N}\), for the glass former network and \(\phi_{nf}:\Omega_{nf}\to\mathbb{R}^{N_{nf}}\), \(N_{nf}\in\mathbb{N}\), for the glass non-former network. Permutation invariance then holds only within the feature vectors \(v_1,\dots,v_{M_f}\) corresponding to the \(M_f\) glass formers and within the \(M_{nf} := M - M_f\) feature vectors \(v_{M_f + 1}, \dots, v_M\) corresponding to the glass non-formers. The resulting representation of \(f\) then has the following form,
\begin{equation}\label{eq: informed func on sets}
    f(\{v_1, \ldots, v_M\}) = \psi\left(\sum_{i=1}^{M_f} \phi_f(v_i), \sum_{i=M_f + 1}^M \phi_{nf}(v_i)\right) \, ,
\end{equation}
where, under slight abuse of notation, we used the same notation \(\psi\) for the downstream function as in \eqref{eq: func on sets}. As glass former network, glass non-former network, and downstream network we use three separate ReLU-FFNNs to approximate the functions \(\phi_f, \phi_{nf}\), and \(\psi\) in~\eqref{eq: informed func on sets}, respectively. Their widths and depths are listed in Table \ref{table: model hyperparameters}. The overall network architecture of our informed model is illustrated in Fig.~\ref{fig: InformedNet}. 

The embedding dimensions \(N_f\) and \(N_{nf}\) are hyperparameters of the glass former and non-former network, respectively. It is shown in \cite{wagstaff2022universal} that in the scalar case, \(d=1\), the choice \(N = M\) in \eqref{eq: func on sets} is a sufficient and necessary condition in order to approximate the function \(f\) arbitrarily well by a neural network whose architecture is given by the right-hand side in \eqref{eq: func on sets}. In the vector-valued case, \(d > 1\), to the best of our knowledge, no non-trivial necessary condition on the embedding dimension is known so far. In~\cite{han2019universal}, the authors prove a sufficient condition in form of an upper bound on the embedding dimension, which, however, is very pessimistic. Based on the results in the one-dimensional case, we choose \(N_f = M_f\) and \(N_{nf} = M_{nf}\) in~\eqref{eq: informed func on sets}.

\subsection{Model Training and Evaluation}
\label{subsec: model training and evaluation}

Recall that we consider three different glass material properties: glass transition temperature \(T_g\), Young's modulus \(E\) at room temperature, and shear modulus \(G\). For each of these properties \Vec{P}, we split the cleaned datasets from Sect.~\ref{subsec: data collection and preparation} further into datasets for training, validation, and testing. We then apply the blind models and the informed model discussed in Sect.~\ref{subsubsec: blind models} and Sect.~\ref{subsubsec: informed model}. 

For data management and visualization, we use the Python libraries \texttt{pandas}~\cite{pandas}, \texttt{Scikit-learn}~\cite{scikit-learn2011}, and \texttt{Matplotlib}~\cite{hunter2007matplotlib}, respectively.
The neural network models are built using the \texttt{PyTorch-Lightning}~\cite{falcon2019pytorchlightning} module.

We describe the data splitting in a bit more detail. For each property \Vec{P}, we apply the following steps. First, we apply the preprocessing pipeline described in Sect.~\ref{subsec: data collection and preparation}. Then, we apply the feature design and selection processes for the blind and the informed models as described in Sect.~\ref{subsubsubsec: blind: feature vectors} and Sect.~\ref{subsubsubsec: informed: feature vectors}, respectively. Next, we split up the resulting cleaned dataset into those glass samples which contain sodium and those which do not contain sodium. From the samples which contain sodium we extract only those binary oxides which consist of \ch{B2O3} as glass former and \ch{Na2O} as glass non-former. The resulting dataset is our \Vec{P}-test set. The other glass samples, which do not contain sodium, are randomly split for each model into a \Vec{P}-training set and a \Vec{P}-validation set using a \(80\%/20\%\) ratio.
The dimensions of the resulting datasets are shown in Table~\ref{table: datasets dimensions}.
We emphasize that sodium as an element is totally absent in the training and validation sets and only present in the test sets. Examining the performance of the trained models on the test sets therefore allows us to properly evaluate their extrapolation power.

\begin{table}[h]\centering
\caption{Dimensions of the training, validation, and test sets.}
\label{table: datasets dimensions}
\ra{1.3}
    \begin{tabular}{@{} p{0.5cm} S[table-format = 4] S[table-format = 3] S[table-format = 3] @{}}          \hline\noalign{\smallskip}
        \Vec{P} & {\#Training samples \ } &  {\ \#Validation samples \ } & {\ \#Test samples}\\ \noalign{\smallskip}\svhline\noalign{\smallskip}
        \(T_g\) & 1385 & 347 & 125 \\
        \(E\) & 415 & 104 & 42 \\
        \(G\) & 477 & 120 & 73  \\ 
        \noalign{\smallskip}\hline\noalign{\smallskip}
    \end{tabular}
\end{table}

We use the \emph{bagging} method from the field of ensemble learning \cite{breiman1996bagging}. For each model setup and each property \Vec{P}, we train 50 models. Their architectures and weight initializations are the same, but each model is trained and validated on a different random \(80\%/20\%\)-split into training set and validation set. We therefore end up with a model ensemble of 50 different models. 

For training, we use the ADAM optimizer with default settings \cite{kingma2014adam} and weight decay of \num{e-5} and train for a maximum of \num{1000} epochs with a batch size of 8. We start with a learning rate of \num{0.001} and multiply it by a factor of \num{0.5} if the model's performance on the validation set in terms of the mean squared error (MSE) does not improve over the course of 50 epochs. Moreover, to avoid overfitting, we use the early stopping criterion and stop training if the model's MSE on the validation set does not improve over the course of 100 epochs.

After training, we apply a post-processing step. For each property \Vec{P}, we discard those models whose predictions for \Vec{P} on the whole test set can be considered to be constant. More specifically, we first compute for each sample in the test set the mean value of the models' predictions for \Vec{P}. Then, we drop those models where the deviation of the predicted property values for all samples in the test set from the respective mean value is less than or equal to the \Vec{P}-duplicate threshold from Table \ref{table: min_max, duplicate_thresholds}. This is in alignment with informed learning since we know a priori that for each property \Vec{P}, not all \ch{Na2O}-\ch{B2O3} glass samples have the same \Vec{P}-value. In terms of the taxonomy in~\cite{rueden2021informed}, we thus use this expert knowledge, which is represented as algebraic equations, i.e., being of non-constant value, and integrate it into our final hypothesis. 

Among the remaining models we compute the mean and the \(95\%\)-confidence interval of the predictions. This yields the final prediction of the model ensemble and quantifies its uncertainty. 
We compare the ensembles' performances quantitatively in terms of their root mean squared errors (RMSE), mean absolute errors (MAE), and maximum errors (MAX) on the respective \Vec{P}-test sets, which are summarized in Table~\ref{table: results and errors}. All blind 32-layer networks yield constant predictions for all three properties and are therefore discarded as non-physical in the post-processing step. Nevertheless, we still record their respective ensemble errors in Table~\ref{table: results and errors} to get a more conclusive picture. However, when talking about the best and worst error values, we \emph{only consider the ensembles of blind models with depths \(L=2,4,8,16\)} and neglect the values of the models with 32 layers.

\begin{figure}
\centering
\includegraphics[scale=.33]{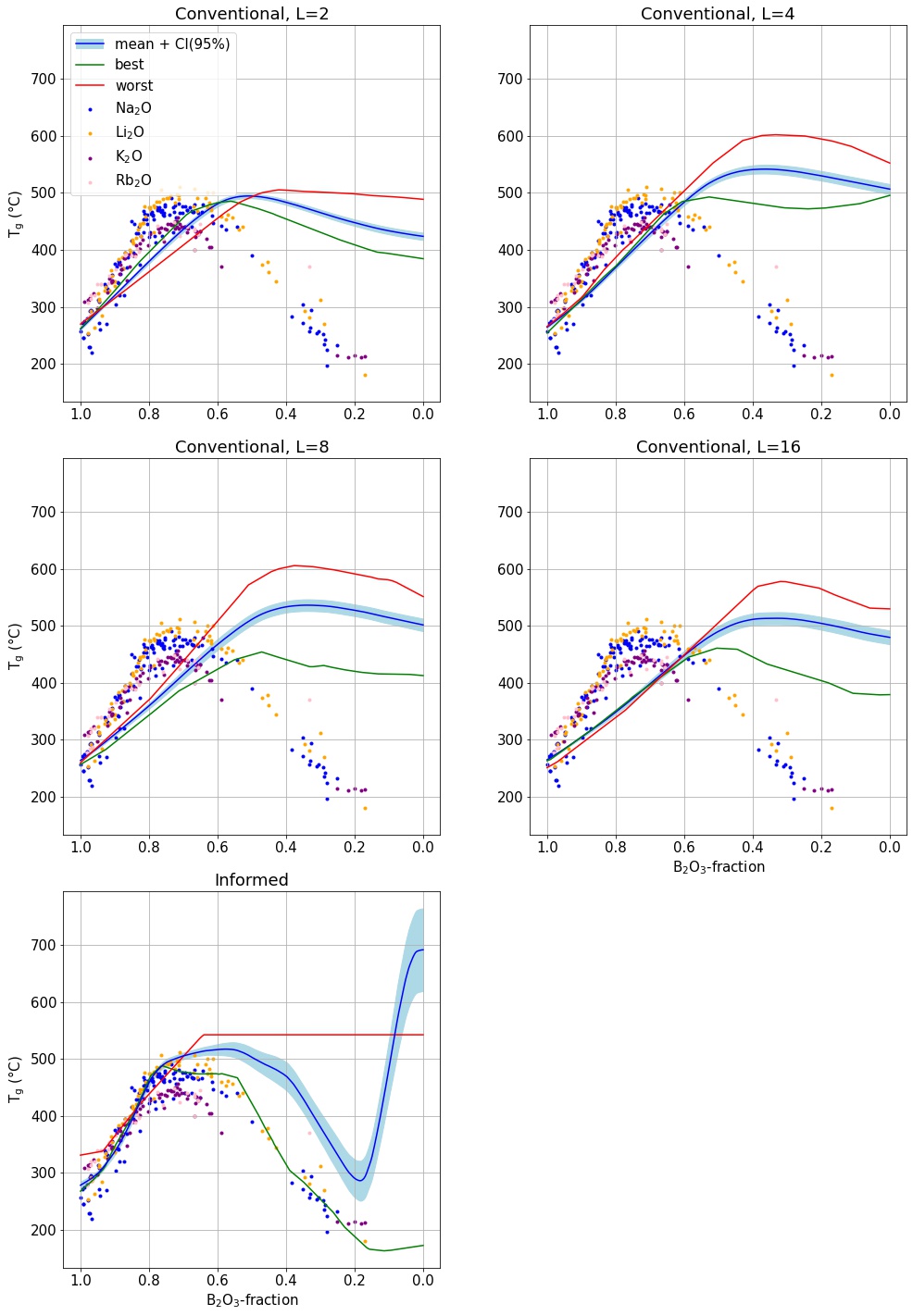}
\caption{\(T_g\)-values of binary alkali borate glasses. Scattered points represent \(T_g\)-values given in the cleaned \(T_g\)-dataset. Solid lines show the predictions for the \(T_g\)-value of \ch{Na2O}-\ch{B2O3} glass samples of the blind and informed models, respectively. The model ensembles' mean curves are shown as blue solid lines with the shaded blue area depicting the \num{95}\% confidence band. The predictions of the best and worst performing models in the ensembles are shown as green and red solid lines, respectively.}
\label{fig: Tg}
\end{figure}

\begin{figure}[h]
\centering
\includegraphics[scale=.33]{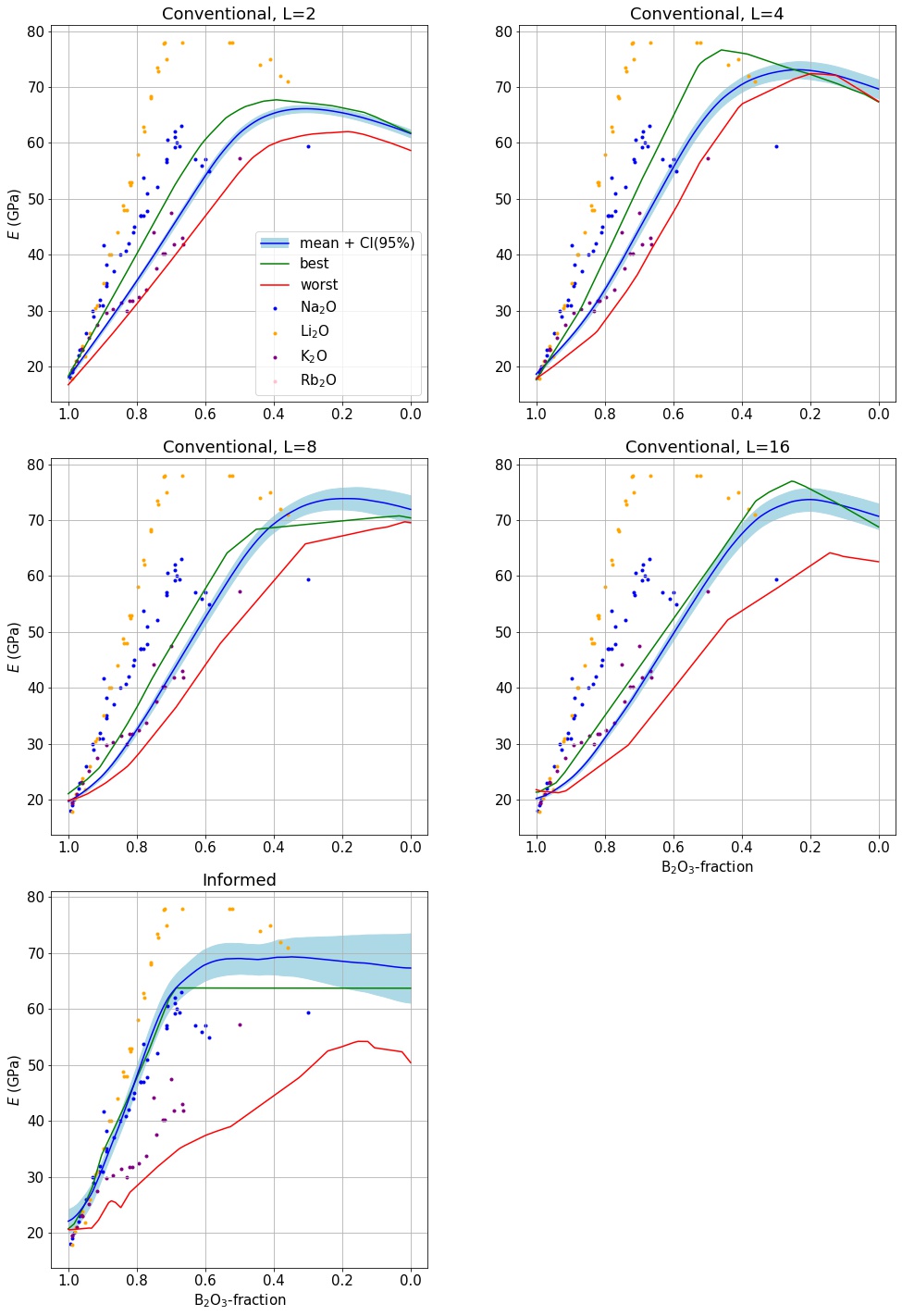}
\caption{\(E\)-values of binary alkali borate glasses. Scattered points represent \(E\)-values given in the cleaned \(E\)-dataset. Solid lines show the predictions for the \(E\)-value of \ch{Na2O}-\ch{B2O3} glass samples of the blind and informed models, respectively. The model ensembles' mean curves are shown as blue solid lines with the shaded blue area depicting the \num{95}\% confidence band. The predictions of the best and worst performing models in the ensembles are shown as green and red solid lines, respectively.}
\label{fig: YoungsModulusRT}
\end{figure}

\begin{figure}[h]
\centering
\includegraphics[scale=.33]{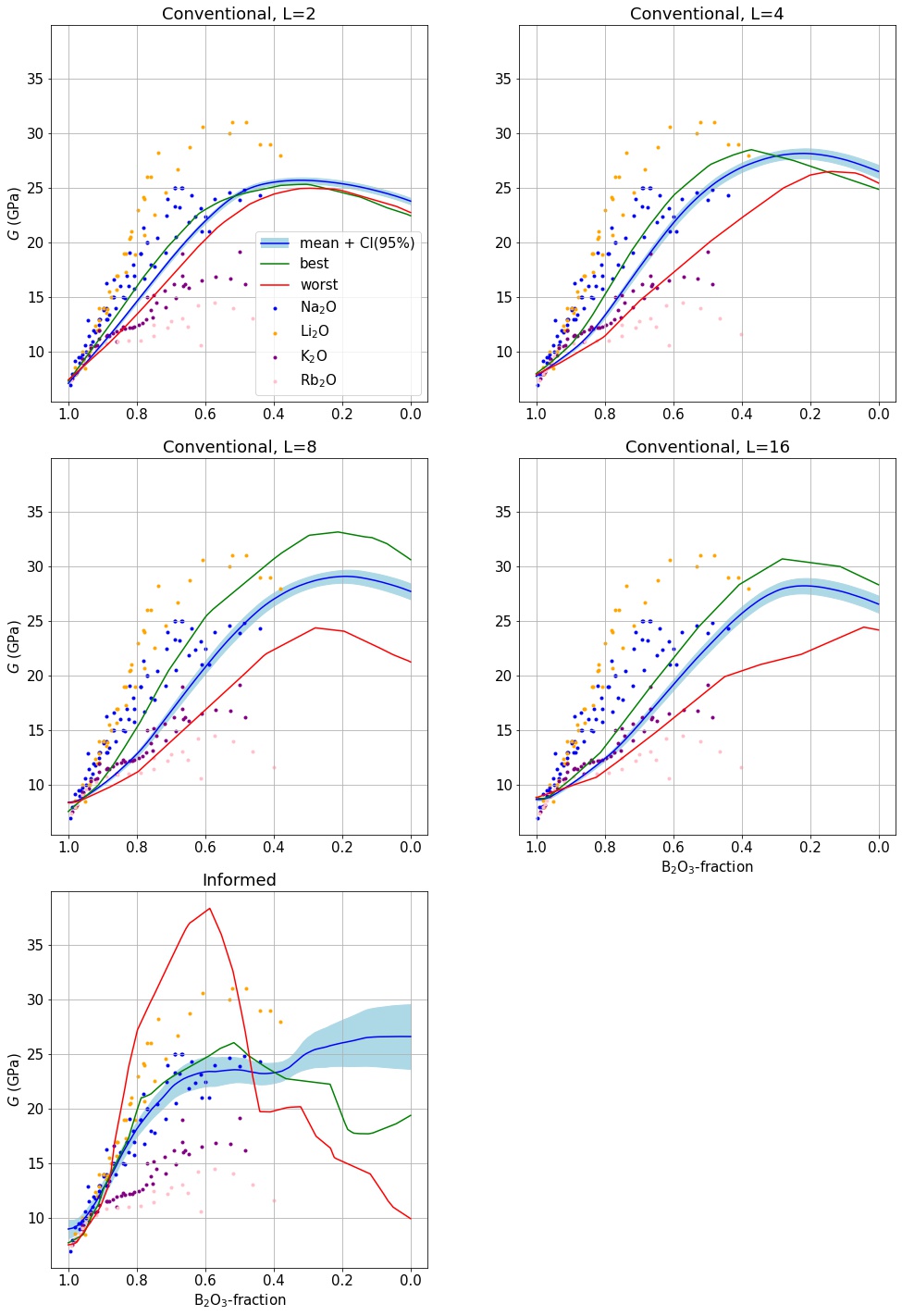}
\caption{\(G\)-values of binary alkali borate glasses. Scattered points represent \(G\)-values given in the cleaned \(G\)-dataset. Solid lines show the predictions for the \(G\)-value of \ch{Na2O}-\ch{B2O3} glass samples of the blind and informed models, respectively. The model ensembles' mean curves are shown as blue solid lines with the shaded blue area depicting the \num{95}\% confidence band. The predictions of the best and worst performing models in the ensembles are shown as green and red solid lines, respectively.}
\label{fig: ShearModulus}
\end{figure}

To also get a qualitative picture of the ensembles' extrapolation performances, we plot the composition-property curves of the ensembles' averaged predictions on the \Vec{P}-test sets in Figs.~\ref{fig: Tg}--\ref{fig: ShearModulus}. Recall that the test sets consist of the \ch{Na2O}-\ch{B2O3} glass samples together with their respective \Vec{P}-values. We also plot the predictions of the best and worst performing model of each \Vec{P}-ensemble in terms of the RMSE on the test set. Moreover, we plot the property values of all available alkali borate glasses, that is, binary glasses which consist of \ch{B2O3} as glass former and \ch{Na2O}, \ch{Li2O}, and \ch{Rb2O} as glass non-former, respectively. It is known that these glass compositions have similar material properties~\cite{feller2019borate}. 

Concerning Figs.~\ref{fig: Tg}--\ref{fig: ShearModulus}, a few remarks are in order. First, there are no \ch{Rb2O}-\ch{B2O3} glass samples available for \(E\). Next, since we only consider binary oxide glasses, knowing the compound fraction of \ch{B2O3} completely determines the compound fraction of the respective alkali oxide as glass non-former as well. Finally, since the blind \(32\)-layer networks are discarded, their predictions are not shown.

\begin{table}[h]\centering
\caption{Results and errors of the model ensembles' averaged predictions on the test sets. Among the blind ensembles with depths \(L=2,4,8,16\), bold blue numbers denote the lowest error values, bold red numbers the highest ones. Bold black numbers denote the lowest error values among all model ensembles. The last column shows the relative improvement in the error when comparing the error value of the informed ensemble to the lowest error value (blue) among the blind ensembles. The blind models with \num{32} layers are not considered in the error analysis as these models only yield constant predictions and are therefore discarded as non-physical.}
\label{table: results and errors}
\ra{1.3}
    \begin{tabular}{@{} p{1.5cm}*{7}{S[table-format=3.2, detect-weight]}P{1.5cm}P{1.5cm} @{}}
        \hline\noalign{\smallskip}
        \multirow{2}{*}{\(T_g\) (\textdegree{}C)} & \multicolumn{6}{c}{{Blind}} &  &  \\ 
        \cline{2-7}
         & {Depth} & {2} & {4} & {8} & {16} & {32} & {\multirow{-2}{*}{\; Informed \;}} & {\multirow{-2}{*}{Rel. improv.}} \\ 
         \noalign{\smallskip}\svhline\noalign{\smallskip}
        RMSE & & \bfseries\color{blue} 86.2 & 111 & \bfseries\color{red} 112 & 109 & 86.5 & \bfseries 62.5 & 27\% \\
        MAE & & \bfseries\color{blue} 63.0 & 77.2 & 81.7 & \bfseries\color{red} 84.4 & 73.6 & \bfseries 44.4 & 30\% \\
        MAX & & \bfseries\color{blue} 265 & \bfseries\color{red} 341 & 338 & 311 & 205 & \bfseries 186 & 30\% \\
        \#Non-const. predictions &  & 50 & 50 & 50 & 48 & 0 & 24 &  \\ 
        \noalign{\smallskip}\hline\noalign{\smallskip}
        \\ 
        \hline\noalign{\smallskip}
        \multirow{2}{*}{\(E\) (GPa)} & \multicolumn{6}{c}{Blind} &  &  \\ \cline{2-7}
         & {Depth} & {2} & {4} & {8} & {16} & {32} & {\multirow{-2}{*}{\; Informed \;}} & {\multirow{-2}{*}{Rel. improv.}}  \\ 
        \noalign{\smallskip}\svhline\noalign{\smallskip}
        RMSE & & \bfseries\color{blue} 9.88 & 10.7 & \bfseries\color{red} 12.0 & 11.6 & 15.6 & \bfseries 4.67 & 53\% \\
        MAE & & \bfseries\color{blue} 8.58 & 9.44 & \bfseries\color{red} 10.7 & 10.3 & 12.5 & \bfseries 3.50 & 59\% \\
        MAX & & \bfseries\color{blue} 16.8 & 18.2 & 19.5 & \bfseries\color{red} 19.8 & 33.7 & \bfseries 12.5 & 26\% \\
        \#Non-const. predictions & & 50 & 50 & 50 & 43 & 0 & 48 &  \\ 
        \noalign{\smallskip}\hline\noalign{\smallskip}
        \\ 
        \hline\noalign{\smallskip}
        \multirow{2}{*}{\(G\) (GPa)} & \multicolumn{6}{c}{Blind} &  &  \\ \cline{2-7}
         & {Depth} & {2} & {4} & {8} & {16} & {32} & {\multirow{-2}{*}{\; Informed \;}} & \multirow{-2}{*}{{Rel. improv.}}  \\ 
        \noalign{\smallskip}\svhline\noalign{\smallskip}
        RMSE & & \bfseries\color{blue} 2.92 & 3.65 & \bfseries\color{red} 4.02 & 3.81 & 5.94 & \bfseries 1.39 & 52\% \\
        MAE & & \bfseries\color{blue} 2.43 & 3.09 & \bfseries\color{red} 3.39 & 3.13 & 5.08 & \bfseries 1.12 & 54\% \\
        MAX & & \bfseries\color{blue} 6.10 & 7.24 & 7.89 & \bfseries\color{red} 8.37 & 12.2 & \bfseries 2.95 & 52\% \\
        \#Non-const. predictions & & 50 & 50 & 50 & 39 & 0 & 48 &  \\ 
        \noalign{\smallskip}\hline\noalign{\smallskip}
    \end{tabular}
\end{table}

\section{Results and Discussion}
\label{sec: results and discussion}

We first compare the models' performances quantitatively in terms of their average errors in Table~\ref{table: results and errors}. Among the blind models, we note that the ensembles of shallow two-layer networks perform best for all three properties in terms of RMSE, MAE, and MAX. Considering the worst performing ensembles, we note that the deeper networks with depths of \num{8} and \num{16} layers perform worst, on average, in terms of almost all three error metrics for all three properties. Only for \(T_g\) in the case of MAX, the ensemble of models with only four layers performs worst. We note that in this case, the ensemble of models with a depth of \num{32} layers actually performs best in terms of MAX. In all other cases, however, the \num{32}-layer network ensembles perform worst for all three properties when compared to the other blind models. We conclude that, in general, increasing network complexity in terms of increasing depth tends to lead to worse performing models.

The number of models with non-constant predictions clearly decays with increasing network depth for all three properties. Whereas the networks with depths of \(2, 4\), and \(8\) layers lead to no constant predictions, the \(16\)-layer networks lead to some constant predictions. There is a steep decay when increasing the number of layers from \(16\) to \(32\), where all models for all properties lead to only constant predictions. A possible explanation for this phenomenon could be that the models' loss landscapes become more and more rugged with increasing network depth yielding constant predictions to be local minima which are hard to escape during the optimization routine. This matches the observation from above that increasing network depth generally tends to lead to worse performing models.

Invoking now the errors of the informed models, we see that they perform best, on average, in terms of all three error metrics for all three properties. They lead to a relative improvement in the errors between 26\% up to 59\%. For \(E\) and \(G\), only two models yield constant predictions, whereas for \(T_g\) more than half of all models do. Again, this could indicate that the loss landscape of the informed networks is much more rugged for \(T_g\) than for the other two properties.

To get a more conclusive qualitative picture of the extrapolation behavior of all models, we take a closer look at Figs.~\ref{fig: Tg}--\ref{fig: ShearModulus}. We observe that the blind networks in terms of the ensembles' means as well as the best and worst performing models are not able to qualitatively capture the trend of the \ch{Na2O}-\ch{B2O3} curves correctly and instead generally deviate from the test points to a large extent. However, the ensembles' predictions for all blind networks seem to be quite close to each other for all three properties. This is reflected by the small width of the confidence band around the mean curves as well as the similar shape of the mean curves and the curves of the best and worst performing models. This indicates that the blind models are robust with respect to training.

The mean curves of the informed model ensembles qualitatively capture the trend of the \ch{Na2O}-\ch{B2O3} curves to a more acceptable degree. This is most noticeable in the cases of \(T_g\) and \(G\) where the mean curves are able to capture the nonlinearity of the respective \ch{Na2O}-\ch{B2O3} curve, which the blind networks are not capable of. For \(E\), the informed model ensemble yields more accurate trajectories than the blind ensembles in the linear regime with \ch{B2O3}-fractions between \num{0.7} and \num{1.0}, but the kink in the \ch{Na2O}-\ch{B2O3} curve at a \ch{B2O3}-fraction of around \num{0.7} is not captured. This could, in parts, be due to the small training and validation sets which are available for \(E\) (see Table~\ref{table: datasets dimensions}) and, in particular, to the lack of \ch{Rb2O}-\ch{B2O3} glass samples which the models could base their predictions on. We explain the latter point in more detail below. 
Nevertheless, whereas the mean curve for \(E\) shows at least a physically reasonable trajectory in the region of low \ch{B2O3}-fractions, where there are no data points of alkali borate glasses available, the mean curves for \(T_g\) and \(G\) show a non-physical incline for glasses of \ch{B2O3}-fractions of less than \num{0.2} and \num{0.4}, respectively.
As a further observation, we note that, in general, for all three properties, the uncertainty of the model ensembles' predictions in terms of the width of the confidence band around the mean curve is much higher than in the blind settings, especially in the regions of low \ch{B2O3}-fractions where there are only few or no alkali borate glass samples available. This indicates less robustness with respect to training the models and is also most noticeably reflected by the large deviation of the worst performing model's curve from the mean curve for all three properties.

As most probable explanation, we suspect these observations to be caused by the choice of our training and test sets. As already indicated in Sect.~\ref{subsec: model training and evaluation}, we note that the curves of all alkali borate glasses show a similar trajectory for all three properties since these glasses have similar material properties. We also note that only \ch{Na2O}-\ch{B2O3} glasses are not present in the training and validation sets. In regions where the other alkali borate glasses are available in the training and validation sets, the models are thus, in principal, able to learn the properties of the \ch{Na2O}-\ch{B2O3} glasses based on the other alkali borate glass samples. In regions where there are many of these samples available and where their property curves are very close to the \ch{Na2O}-\ch{B2O3} curve, the informed models' predictions thus tend to be quite accurate.
In regions where only few or no alkali borate glasses are available in the training and validation sets, the models are prone to base their predictions on other spurious or noisy features. This leads to non-physical predictions with a high uncertainty. This phenomenon is amplified by a large feature set and is therefore pronounced to a much higher degree in the informed setting than in the blind one.

In summary, the informed model shows, on average in the ensemble setting, clear superior performance than all considered blind (uninformed) models in extrapolating the property curves of \ch{Na2O}-\ch{B2O3} binary glasses for all three properties \(T_g\), \(E\), and \(G\). This is in terms of quantitative error measurements on the test sets as well as in the qualitative approximation of the property curves. 

Finally, we emphasize the importance of the ensemble setting. Whereas single models might yield bad predictions, averaging multiple trained models, as we observe in our specific use case, often yields good approximations of the target quantities~\cite{rokach2010ensemble}.

\section{Conclusion and Outlook}
\label{sec: conclusion and outlook}

In this paper, we presented an informed neural network approach for the prediction of three material properties of binary oxide glasses, that is, glass transition temperature \(T_g\), Young's modulus \(E\) (at room temperature), and shear modulus \(G\). We compared this approach to five different blind (uninformed) models for all three properties and demonstrated its superior average extrapolation power when applied in an ensemble setting to alkali borate glass samples which contain sodium as previously unseen element. 

In terms of the taxonomy of informed machine learning introduced in~\cite{rueden2021informed}, we integrated prior knowledge into our learning pipeline at four major points. We integrated scientific knowledge, represented as a weighted graph, knowledge graph, and spatial invariance in the training data and in the hypothesis set, respectively. Moreover, we integrated expert knowledge, represented as algebraic equations, into the final hypothesis.

Our informed neural network model could be improved in various ways. First, the list of chemical and physical element features could be extended. Second, instead of classifying glass oxides into formers and non-formers, we could follow the refined classification into formers, modifiers, and intermediates and treat these three classes by three separate neural networks. Third, in this paper, we did not tune any of the models' hyperparameters. A thorough hyperparameter study probably leads to improved model performance. Finally, by relying on further expert knowledge, we could potentially filter out even more predicted property curves in the post-processing step than just constant predictions. This might improve the final predictions even further. 

Our results show that our informed neural network model is capable of meaningfully extrapolating various properties of binary glass samples with previously unseen compounds. As a next step, we plan to scale up our approach in order to make it applicable to oxide glass samples with three or more compounds. We also plan to make it more universal, such that it can accurately predict more material properties.

\begin{acknowledgement}
This work was supported in part by the BMBF-project 05M2AAA MaGriDo (Mathematics for Machine Learning Methods for Graph-Based Data with Integrated Domain Knowledge), by the Fraunhofer Cluster of Excellence Cognitive Internet Technologies, and by the Deutsche Forschungsgemeinschaft (DFG, German Research
Foundation) via project 390685813 - GZ 2047/1 - Hausdorff Center for Mathematics (HCM).
\end{acknowledgement}

\clearpage

\begin{table}[h]\centering
\ra{1.3}
\caption{Chemical and physical properties extracted from the \texttt{mendeleev} library which are used for the correlation study described in Sect.~\ref{subsubsubsec: informed: feature vectors}. Properties that are marked with {\xmark} are dropped before the correlation study as they are not available for all respective \Vec{P}-elements. Properties that are marked with {\cmark} are not highly correlated among each other and are used as final features. All unmarked properties are dropped due to too high correlation with other properties. We refer to \cite{mendeleev2014} and the references therein for detailed explanations of the properties.}
\label{table: mendeleev properties}
    \begin{tabular}{@{} p{6cm}p{1cm}p{1cm}p{0.3cm} @{}} \hline\noalign{\smallskip}
    Properties from \texttt{mendeleev} & \(T_g\) & \(E\) & \(G\) \\ 
    \noalign{\smallskip}\svhline\noalign{\smallskip}
    Atomic number & \cmark & \cmark & \cmark \\
    Atomic radius & \cmark & \cmark & \cmark \\
    Atomic radius by Rahm et al. & \cmark & \cmark & \cmark \\
    Atomic volume & \cmark & \cmark & \cmark \\
    Atomic weight &  &  &  \\
    Boiling temperature & \cmark & \cmark & \cmark \\
    \(C_6\) dispersion coefficient by Gould and Bučko & \cmark & \cmark & \cmark \\
    Covalent radius by Cordero et al. & \cmark &  &  \\
    Single bond covalent radius by Pyykko et al. &  &  &  \\
    Double bond covalent radius by Pyykko et al. &  &  &  \\
    Density & \cmark & \cmark & \cmark \\
    Dipole polarizability &  &  &  \\
    Electron affinity & \cmark & \cmark & \cmark \\
    Electron affinity in the Allen scale & \xmark & \cmark & \cmark \\
    Electron affinity in the Ghosh scale & \cmark & \cmark & \cmark \\
    Electron affinity in the Pauling scale & \cmark &  &  \\
    Glawe’s number & \cmark & \cmark & \cmark \\
    Group in periodic table & \cmark & \cmark & \cmark \\
    Heat of formation & \cmark & \cmark & \cmark \\
    First ionization energy & \cmark & \cmark & \cmark \\
    Lattice constant & \cmark & \cmark & \cmark \\
    Maximum coordination number & \cmark & \cmark & \cmark \\
    Maximum oxidation state & \cmark & \cmark & \cmark \\
    Melting temperature & \cmark &  &  \\
    Mendeleev's number &  &  &  \\
    Minimum coordination number & \cmark & \cmark & \cmark \\
    Minimum oxidation state & \cmark & \cmark & \cmark \\
    Period in periodic table & \cmark & \cmark & \cmark \\
    Pettifor scale &  &  &  \\
    Index to chemical series & \cmark & \cmark & \cmark \\
    Number of valence electrons & \cmark & \cmark & \cmark \\
    Van der Waals radius & \cmark & \cmark & \cmark \\
    Van der Waals radius according to Alvarez & \cmark & \cmark & \cmark \\
    Van der Waals radius according to Batsanov &  &  &  \\
    Van der Waals radius from the MM3 FF &  &  &  \\
    Van der Waals radius from the UFF & \cmark & \cmark & \cmark \\ 
    \noalign{\smallskip}\svhline\noalign{\smallskip}
    {Number \(d\) of all features \hspace{3cm} \linebreak (including mole atomic fractions)} & 28 & 26 & 26 \\ 
    \noalign{\smallskip}\hline\noalign{\smallskip}
    \end{tabular}
\end{table}

\clearpage 

\bibliographystyle{spmpsci}
\bibliography{references}

\end{document}